\documentclass[pr,twocolumn,
showkeys,showpacs,nofootinbib,byrevtex]{revtex4-1}
\usepackage{amsmath,amsfonts,amssymb,amscd,amsxtra,amsthm}
\usepackage{graphicx}
\usepackage{bm}
\usepackage{epstopdf}
\usepackage{subfigure}
\usepackage{ulem}
\begin{document}

\title{Investigating convergence of the reaction  $\gamma
p\to\pi^\pm\Delta$ and  tensor meson $a_2$ exchange at high
energy}

\author{Byung-Geel Yu}
\email[E-mail: ]{bgyu@kau.ac.kr} \affiliation{Research Institute
of Basic Sciences, Korea Aerospace University, Goyang, 412-791,
Republic of Korea}
\author{Kook-Jin Kong}
\email[E-mail: ]{kong@kau.ac.kr} \affiliation{Research Institute
of Basic Sciences, Korea Aerospace University, Goyang, 412-791,
Republic of Korea}
\date{\today}

\begin{abstract}
A Regge approach to the reaction processes $\gamma
p\to\pi^-\Delta^{++}$ and $\gamma p\to\pi^+\Delta^0$ is presented
for the description of existing data up to $E_\gamma= 16$ GeV. The
model consists of the $t$-channel $\pi(139)+\rho(775)+a_2(1320)$
exchanges which are reggeized from the relevant Born amplitude.
Discussion is given on the minimal gauge prescription for the
$\pi$ exchange to render convergent  the divergence of the
$u$-channel $\Delta$-pole in the former process. A new Lagrangian
is constructed for the $a_2N\Delta$ coupling in this work and
applied to the process for the first time with the coupling
constant deduced from the duality plus vector dominance. It is
shown that, while the $\pi$ exchange dominates over the process,
the role of the $a_2$ exchange is crucial rather than the $\rho$
in reproducing the cross sections for total, differential, and
photon polarization asymmetry to agree with data at high energy.
\end{abstract}


\maketitle

Photoproduction of $\Delta$-baryon is an example of the hadron
reaction to study the formation of a composite particle from the
$\pi N$ interaction. Efforts in existing theories and experiments
on this reaction process  have been concentrated mainly on
understanding of the dynamical and static properties of
$\Delta$-baryon around the $\Delta$-resonance region
\cite{nacher,ripani,pascal}.
But, it could also be of value to question how the production
mechanism would proceed over the region, and it is, therefore, a
challenging issue to establish a model for the $\gamma
p\to\pi^-\Delta^{++}$ process which is valid at high energy,
because the $\Delta$-propagation in the process would be highly
divergent there.

In experiments evidences are clear for the formation of the
$\Delta$-baryon through the reaction $\gamma N\to \pi\pi N$ in the
photon energy below the threshold of $\rho N$ production.
At high energies the reaction cross sections for the $\gamma
p\to\pi^-\Delta^{++}$ process were measured for the differential,
spin observables for the photon beam and density matrix elements
up to photon energy $E_\gamma=16$ GeV
\cite{Boyarski:1969dw,Bingham:70,quinn79}.
Only recently the total and differential cross sections for
$\gamma p\to\pi^-\Delta^{++}$ were obtained from threshold up to
$\sqrt{s}=2.6$ GeV at the ELSA \cite{wu2005}.
In particular, the total cross section for $\gamma
p\to\pi^-\Delta^{++}$ at the ELSA shows a sharp peak of
$\sigma_{max}\approx$ 70 $\mu$b in size around $E_\gamma\approx$
1.6 GeV with the steep decrease following over the resonance
region, and, hence, exhibits a typical feature of the
nondiffractive two-body process.

In this work  we will analyze the production mechanism of the
reactions $\gamma p\to \pi^-\Delta^{++}$ and $\gamma p\to
\pi^+\Delta^0$ at high energies with a focus on the convergence of
the reaction processes there. From a theoretical point of view
only the $\pi$ exchange in the $t$-channel peripheral subprocess
is expected to dominate at high energies and small momentum
transfer. Hence the production amplitude should be ${\cal
M}\propto {q\cdot\epsilon/( t-m_\pi^2)}$, where $\epsilon$ is
photon polarization, and $q$ and $m_\pi$ are pion momentum and
mass. However, since the $t$-channel $\pi$ exchange itself is not
gauge invariant, an extension of the production amplitude is
needed for gauge invariance. Furthermore this should be a
specialized one for the extended amplitude has to be convergent at
high energy, even if it includes highly divergent $\Delta$
propagation for gauge invariance. For this requirement a
theoretical speculation was suggested in Ref. \cite{stichel} that
the amplitude, thus extended, contain only the charge couplings of
the $s$-channel proton-pole and $u$-channel $\Delta$-pole coupling
to photon field in the $\gamma p\to\pi^-\Delta^{++}$ process. In
other words, the transverse components in these $s$ and
$u$-channel poles  should be removed in order for the convergence
of the process to be ensured at high energy. This leads to the
so-called the minimal gauge prescription in the sense that the
proton-pole and  $\Delta$-pole terms  are minimally introduced for
gauge invariance of the $t$-channel $\pi$ exchange. Moreover, such
a scheme seems to be reasonable because the higher multipoles of
the $\Delta$-baryon and proton electromagnetic moments are defined
uniquely in the static limit and such a uniqueness can no longer
be valid at high energy.

Application of the minimal gauge condition is found in Ref.
\cite{clark1} in which case the dynamics of $\gamma N\to\pi^\pm
\Delta$ was investigated in the kinematical region, $-t\leq
m_\pi^2$ and $s\to\infty$ at forward angles. A more qualitative
analysis of the reaction was made in Ref. \cite{goldstein} by
using the $\pi+b_1+\rho+a_2$ Regge-pole exchanges in the
$t$-channel helicity amplitude with their residues and cuts
considered to fit to data.

On the other hand, we note that the tensor-meson $a_2$ plays the
role to significantly improve the cross sections for the
differential and spin polarizations in the $\gamma N\to\pi^\pm N$
process at high energy \cite{bgyu-pi}. The significance of such a
higher-spin interaction is confirmed in the cases of $\gamma p\to
K^+\Lambda\,(\Sigma^0)$ as well by the role of the $K_2^*$
\cite{bgyu-kaon}. Nevertheless, there are no attempts at present,
however, to investigate the role of the tensor meson $a_2$ in the
$\pi\Delta$ photoproduction utilizing the effective Lagrangian
except for the case of the Regge-pole fit to data discussed above.

In this work our interest is to  construct  a model for the
$\gamma p\to\pi^-\Delta^{++}$ process  at high energy where the
reaction cross sections can be described without either fit
parameters or any counter terms included in $ad$ $hoc$ fashion.
For this purpose we consider to incorporate  the two basic
ingredients with the model, i.e., the minimal gauge-invariance and
the role of the spin-2 tensor-meson exchange.

\begin{figure}[]
    \centering
    \includegraphics[width=0.6\hsize,angle=0]{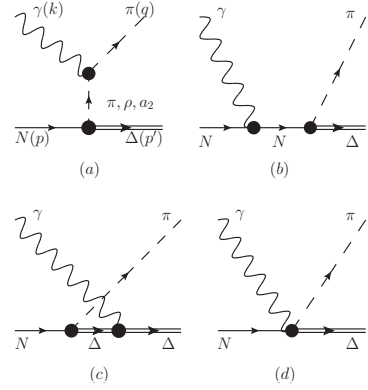}
    \caption{
    Feynman diagrams for the $\gamma N \to\pi \Delta$
    process. $\pi$ exchange in the $t$-channel (a), proton-pole
    in the $s$-channel (b), $\Delta$-pole in the $u$-channel (c), and
    the contact term (d) are the basic ingredients for gauge
    invariance of the reaction. Exchanges of $\rho$ and $a_2$  are
    gauge-invariant themselves in the $t$-channel (a).
    } \label{fig1}
\end{figure}

For a heuristic introduction of the minimal gauge prescription,
let us begin with the Born amplitude for the process,
\begin{eqnarray}\label{process}
\gamma(k)+p(p)\to\pi^\pm(q)+\Delta(p'),\label{reac1}
\end{eqnarray}
where the production amplitude for the $\gamma p\to
\pi^-\Delta^{++}$ process consists of the proton-pole in the
$s$-channel, the $\Delta^{++}$-pole in the $u$-channel,  and the
$\pi^-$ exchange in the $t$-channel to respect the charge
conservation, $e_N-e_\pi-e_{\Delta}=0$, with the charges of
nucleon, $\Delta$, and $\pi$ denoted by $e_N$, $e_\Delta$, and
$e_\pi$, respectively. These are summarized in diagrams in Fig.
\ref{fig1}.

With the contact term further  the reggeized $\pi$ exchange which
is gauge invariant is, thus, given by \cite{bgyu-pi}
\begin{eqnarray}\label{pi-regge}
&&{\cal
M}_{\pi}=\left[{M}_{s(N)}+{M}_{u(\Delta)}+{M}_{t(\pi)}+M_c\right]\nonumber\\
&&\hspace{1.7cm}\times(t-m_\pi^2){\cal R}^\pi(s,t),
\end{eqnarray}
where
\begin{eqnarray} &&i{M}_{s(N)}=\frac{f_{\pi
N\Delta}}{m_{\pi}}
\bar{u}_{\nu}(p')q^\nu\frac{(\rlap{/}{p}+\rlap{/}{k}
+M_{N})}{s-M^{2}_{N}}e_N \rlap{/}{\epsilon}
\,u(p),\label{amp-p}\\
&&i{M}_{u(\Delta)}= -\frac{f_{\pi N\Delta}}{m_{\pi}}
\bar{u}_{\nu}(p')e_\Delta(g^{\nu\alpha}\rlap{/}\epsilon-\epsilon^\nu\gamma^\alpha)\nonumber\\
&&\hspace{1.5cm}\times\frac{
(\rlap{/}{p'}-\rlap{/}{k}+M_{\Delta})}
{u-M^{2}_{\Delta}}\Pi^\Delta_{\alpha\beta}(p'-k)q^{\beta}u(p),\label{amp1-u}\\
&&i{M}_{t(\pi)}= \frac{f_{\pi N\Delta}}{m_{\pi}}\bar{u}_{\nu}(p')
e_{\pi}\frac{(2q-k)\cdot \epsilon }{t-m^{2}_{\pi}}(q-k)^\nu
u(p),\label{amp2}\\
&&i{M}_c=-e_\pi\frac{  f_{\pi N\Delta}}{m_\pi}\
\bar{u}_\nu(p')\epsilon^{\nu} u(p)\,,
\end{eqnarray}
with  the masses of nucleon, $\Delta$, and $\pi$ denoted by $M_N$,
$M_\Delta$, $m_\pi$, respectively. Here, the off-shell effect in
the ${\pi N\Delta}$ vertex is neglected for simplicity.

For the charge coupling of the $\gamma\Delta\Delta$ vertex we use
\begin{eqnarray}
\label{vertex1}
    e_\Delta\epsilon_\mu\Gamma_{\gamma\Delta\Delta}^{\mu\nu\alpha}=
    -e_\Delta(g^{\nu\alpha}\rlap{/}\epsilon-\epsilon^\nu\gamma^\alpha
    -\gamma^\nu \epsilon^\alpha +\gamma^\nu\rlap{/}\epsilon
    \gamma^\alpha),
\end{eqnarray}
which is obtained from the Ward identity at the
$\gamma\Delta\Delta$ vertex
\begin{eqnarray}\label{ward}
\label{gamma1}
k_\mu\Gamma^{\mu\nu\alpha}_{\gamma\Delta\Delta}(p,k)=
(D^{\nu\alpha})^{-1}(p+k) -(D^{\nu\alpha})^{-1}(p),
\end{eqnarray}
by using the propagator,
\begin{eqnarray}\label{propa1}
    D^{\alpha\beta}(p)=\frac{(\rlap{/}{p}+M_\Delta)\Pi_\Delta^{\alpha\beta}(p)}
    {p^2-M_\Delta^2}\,,
\end{eqnarray}
and the spin projection for the spin-3/2 $\Delta$ baryon,
\begin{eqnarray}\label{propa2}
    \Pi_\Delta^{\alpha\beta}(p)=-g^{\alpha\beta}+\frac{1}{3}\gamma^{\alpha}\gamma^{\beta}
    +\frac{\gamma^{\alpha}p^{\beta}-\gamma^{\beta}p^{\alpha}}{3M_\Delta}
    +\frac{2p^{\alpha}p^{\beta}}{3M^{2}_\Delta}\,.\hspace{0.5cm}
\end{eqnarray}
The $t$-channel Regge pole in Eq. (\ref{pi-regge}) is given by
\begin{eqnarray}\label{regge3}
{\cal R}^\varphi
=\frac{\pi\alpha'_\varphi}{\Gamma(\alpha_\varphi(t)+1-J)}
\frac{\mbox{phase}}{\sin\pi\alpha_\varphi(t)}
\left(\frac{s}{s_0}\right)^{\alpha_\varphi(t)-J},\ \
\end{eqnarray}
written collectively for $\varphi(=\pi,\,\rho,\, a_2)$-meson of
spin $J$ with $s_0=$ 1 GeV$^2$. For the phase of the Regge pole
the canonical form of ${1\over 2}((-1)^J+
e^{-i\pi\alpha_\varphi(t)})$ is generally assigned to the exchange
non-degenerate meson \cite{guidal}. In the case of the
exchange-degenerate (EXD) trajectories $\pi$-$b_1$ and
$\rho$-$a_2$ pairs the determination of the phases  will be
discussed later.

\subsection*{The minimal gauge prescription for $\pi$ exchange}

As discussed above the reaction amplitude converging at high
energy should couple to the Coulomb component of  photon field,
and such a condition should be taken into account in the gauge
invariant amplitude in Eq. (\ref{pi-regge}).

We recall that the $u$-channel $\Delta$-pole as well as the
$s$-channel proton-pole term is introduced merely to preserve
gauge invariance for the $t$-channel $\pi$ exchange at high
energy, as discussed in Ref. \cite{stichel}. We, then, consider
only the charge couplings of  the $s$-, and $u$-channel amplitudes
that are indispensable to restore gauge invariance, but remove all
the transverse ones which are redundant by gauge invariance.
In the  proton-pole term in Eq. (\ref{amp-p}), for instance, only
the $2p\cdot\epsilon$ term has to be there, whereas such a gauge
invariant term, $\rlap{/}k \rlap{/}\epsilon$ from the
$(\rlap{/}p+\rlap{/}k+M_N)\rlap{/}\epsilon$ as well as from the
magnetic moment term is redundant and, hence, excluded.

Similarly, in the case of the $u$-channel amplitude in Eq.
(\ref{amp1-u}) which is written in a fully expended form as
\begin{eqnarray}\label{uch1}
&&i{M}_{u(\Delta)}\nonumber\\
&&=e_\Delta \frac{f_{\pi N\Delta}}{m_{\pi}}
\bar{u}^{\nu}(p')\biggl[ q_\nu (2\epsilon \cdot p'
-\rlap{/}{\epsilon}\,\rlap{/}{k}) +\frac{2}{3}
(k_\nu \rlap{/}{\epsilon} -\epsilon_\nu \rlap{/}{k})\,\rlap{/}{q}\, \nonumber\\
&&+\frac{2}{3M_\Delta}
(k_\nu \rlap{/}{\epsilon} -\epsilon_\nu \rlap{/}{k})\,(p'-k)\cdot q\, \nonumber\\
&&-\frac{1}{3M_\Delta} (2k_\nu \,p'\cdot \epsilon -2\epsilon_\nu
\,p'\cdot k
-k_\nu \rlap{/}{\epsilon}\,\rlap{/}{k})\,\rlap{/}{q}\, \nonumber\\
&&+\frac{2}{3M_\Delta^2} (2k_\nu\, p'\cdot \epsilon -2\epsilon_\nu
\,p'\cdot k -k_\nu \rlap{/}{\epsilon}\,\rlap{/}{k})\,(p'-k)\cdot
q\biggr]u(p)\nonumber\\
&&\times{1\over u-M^2_\Delta}\,,
\end{eqnarray}
only the first term proportional to $2\epsilon\cdot p'$ is not
invariant and should be reserved, whereas all the others are
excluded by redundancy. It should be noted that this
simplification is possible only when  the $M_{u(\Delta)}$ term can
be written in such an antisymmetric form as in Eq. (\ref{uch1})
with respect to photon polarization $\epsilon$ and momentum $k$,
and this can only be done by the benefit of the second term
$-\epsilon^\nu\gamma^\alpha$ in the $\gamma\Delta\Delta$ coupling
in Eq. (\ref{amp1-u}). This signifies the validity of the Ward
identity Eq. (\ref{ward}) at the photon coupling vertex.

The invariant amplitude
${M}_{s(N)}+{M}_{u(\Delta)}+{M}_{t(\pi)}+M_c$ in Eq.
(\ref{pi-regge}) in this minimal gauge is, therefore, written as
\cite{stichel,clark1}
\begin{eqnarray}\label{min}
&&\frac{f_{\pi N\Delta}}{m_{\pi}}
\bar{u}_{\nu}(p')\biggl[q^\nu\frac{2p\cdot\epsilon}{s-M^{2}_{N}}e_N
%
+e_\Delta \frac{2p'\cdot\epsilon}
{u-M^{2}_{\Delta}}q^{\nu}\nonumber\\&&\hspace{3cm}
+ e_{\pi}\frac{2q\cdot \epsilon }{t-m^{2}_{\pi}}(q-k)^\nu\biggr]
u(p)+iM_c\,.\ \
\end{eqnarray}
In the numerical analysis, the effect of the minimal gauge on the
cross section will be examined as the photon energy increases up
to $16$ GeV.

The value for the coupling constant $f_{\pi^-p\Delta^{++}}$ is
scattered in various reactions, i.e., from the quark model
prediction $f_{\pi N\Delta}={6\sqrt{2}\over5}f_{\pi NN} \approx
1.7$ with $f_{\pi p p}=1$ for the NN interaction \cite{gebrown} to
$f_{\pi^-p\Delta^{++}}= 2.16$ from the decay width
$\Gamma_{\Delta\to\pi N}=120$ MeV for the $\pi\Delta$
photoproduction \cite{nam-yu}. We consider the one within the
range of $1.7\leq f_{\pi N\Delta}\leq 2.16$ that is better to
agree with experimental data.

\subsection*{$\rho$ and $a_2$ exchanges}

The reggeized $\rho$ exchange  in the $t$-channel is expressed as
\cite{nam-yu},
    \begin{eqnarray}\label{amp2}
    &&i{\cal M}_{\rho}=-i{g_{\gamma{\pi}\rho}\over m_0}\frac{f_{\rho N\Delta}}{m_{\rho}}
    \epsilon^{\alpha\beta\lambda\sigma}
    \epsilon_{\alpha}k_{\beta}q_{\lambda}
    \nonumber\\&&\hspace{1.5cm}\times
    \bar{u}^{\nu}(p')
    \left(\rlap{/}Q g_{\nu\sigma}-Q_{\nu}\gamma_{\sigma}
    \right) \gamma_5 u(p){\cal R}^\rho(s,t),\ \
    \end{eqnarray}
where we use the interaction Lagrangian
\begin{eqnarray}\label{rho}
{\cal L}_{\rho N\Delta}=i{f_{\rho N\Delta}\over m_\rho}
\bar{\Delta}_\nu\gamma_\mu\gamma_5
N(\partial^\nu\rho^\mu-\partial^\mu\rho^\nu)+{\rm h.c.},
\end{eqnarray}
for the $\rho N\Delta$ coupling and neglect other nonleading
terms.
Similar to $\pi N\Delta$, the coupling constant $f_{\rho N\Delta}$
is determined from the relation $f_{\rho
N\Delta}$=${6\sqrt{2}\over 5}f_{\rho NN}$$\approx$ 8.57, where $
{f_{\rho NN}\over m_\rho}=g_{\rho
NN}{(1+\kappa_\rho)\over2M}\approx 5.05/ m_\rho$ is estimated from
the one-boson-exchange in the NN potential \cite{gebrown} by using
$g_{\rho NN}=2.6$ and $\kappa_\rho=3.7$ \cite{workman,bgyu-rho}.
The radiative decay constant is determined from the decay width
$\Gamma_{\rho\to\gamma\pi}$= 67.1 keV and we obtain
$g_{\gamma\rho\pi}$ = $\pm 0.224$, by using the Lagrangian,
\begin{eqnarray}
{\cal L}_{\gamma\pi \rho}=-\frac{g_{\gamma
\pi\rho}}{m_0}\epsilon^{\mu\nu\alpha\beta}\partial_{\alpha}A_{\beta}\partial_\mu\rho_\nu
\pi +\mathrm{h.c.}\,.
\end{eqnarray}

It is expected that the exchange of tensor meson $a_2(1320)$ of
spin-2 plays a role at high energy from the previous studies of
the charged pion photoproduction \cite{clark1,goldstein,bgyu-pi}.
However, no information is available at present either for the
interaction Lagrangian or for the coupling constants of the $a_2
N\Delta$ interaction. In this work we construct a new Lagrangian
for the $a_2N\Delta$ coupling for application.

By considering parities and spins, but neglecting the off-shell
effect in the meson-$\Delta$ coupling for simplicity, we write the
Lagrangian for the  $a_2N\Delta$ coupling as\footnote{ The
interaction form of the Lagrangian in Eq. (\ref{a2}), of course,
might not be unique but one of the possible couplings between
$a_2$ and $N\Delta$ baryon transition-current, and one could also
consider the interaction of form
\begin{eqnarray}
{\cal L}_{a_2 N\Delta}={f_{a_2 N\Delta}}
\bar{\Delta}_\nu\gamma_\mu\gamma_5 Na^{\mu\nu}_2 \nonumber
\end{eqnarray}
by replacing the $\rho^{\nu\mu}$ with $a_2^{\nu\mu}$ in Eq.
(\ref{rho}), for instance, though unnatural to identify the
$\rho^{\nu\mu}$ with $a_2^{\nu\mu}$. In this case, however,
he(she) cannot use the identity in Eq. (\ref{id}) to determine the
$f_{a_2N\Delta}$, but has to consider it as a parameter to fit to
data, because of the different mass dimensions between the
$f_{\rho N\Delta}$ and $f_{a_2 N\Delta}$.}
\begin{eqnarray}\label{a2}
&&{\cal L}_{a_2N\Delta}=i{f_{a_2N\Delta}\over
m_{a_2}}\bar\Delta^\lambda\left(g_{\lambda\mu}\overleftrightarrow{\partial_\nu}
+g_{\lambda\nu}\overleftrightarrow{\partial_\mu}\right) \gamma_5 N
{a_2}^{\mu\nu}\nonumber\\&&\hspace{1.5cm}+\mathrm{h.c.}\,,
\end{eqnarray}
in favor of using the identity to determine the $f_{a_2N\Delta}$
coupling constant,
\begin{eqnarray}\label{id}
{f_{a_2 N\Delta}\over m_{a_2}}=-3{f_{\rho N\Delta}\over
m_{\rho}}\,,
\end{eqnarray}
which follows the duality and vector meson dominance
\cite{goldstein}. Here
$\overleftrightarrow{\partial_\nu}=(\overrightarrow{\partial_\nu}-\overleftarrow{\partial_\nu})/2$.

The $\gamma\pi a_2$ coupling was derived in Ref. \cite{giacosa}
and the interaction Lagrangian is  given by
\begin{eqnarray}
{\cal L}_{\gamma\pi a_2}=-i\frac{g_{\gamma \pi
a_2}}{m^2_0}\tilde{F}_{\alpha\beta}(\partial^\alpha
a_2^{\beta\rho}-\partial^\beta a_2^{\alpha\rho})
\partial_\rho \pi +\mathrm{h.c.}\,,\
\end{eqnarray}
where
$\tilde{F}_{\alpha\beta}={1\over2}\epsilon_{\mu\nu\alpha\beta}F^{\mu\nu}$
is the pseudotensor of photon field. The decay of the
$a_2\to\pi\gamma$ is reported to be
$\Gamma_{a_2^\pm\to\pi^\pm\gamma}=(311\pm25)$ keV in the Particle
Data Group (PDG), which gives  $g_{\gamma\pi a_2}=\pm0.276$.

As the only unknown coupling constant $f_{a_2N\Delta}$ is
determined, once the $f_{\rho N\Delta}$ is given, there are,
therefore, no free parameters in the present calculation. We
choose the signs of the coupling constants $g_{\gamma\pi\rho}$ and
$g_{\gamma\pi a_2}$ to agree with photon polarization asymmetry
$\Sigma$ of $\gamma p\to \pi^-\Delta^{++}$ and $\gamma
p\to\pi^+\Delta^0$.

The reggeized  $a_2$  exchange is given by
\begin{eqnarray}
\label{amp4} && i{\cal M}_{a_2}= -i\frac{2g_{\gamma\pi
a_2}}{m_0^2}{f_{a_2 N\Delta}\over m_{a_2}}
\,\epsilon^{\alpha\beta\mu\lambda}\epsilon_\mu k_\lambda Q_\alpha
q_\rho \nonumber\\
&&\hspace{1cm}\times \Pi_{a_2}^{\beta\rho;\sigma\xi}(Q)
\bar{u}^{\nu}(p')(g_{\nu\sigma}P_\xi+g_{\nu\xi}P_\sigma)\gamma_5
u(p)\nonumber\\
&&\hspace{1cm}\times{\cal R}^{a_2}(s,t),
\end{eqnarray}
where $P=(p+p')/2$ and the spin-2 projection is
\begin{eqnarray}
\Pi^{\beta\rho;\sigma\xi}_{a_2}(Q)={1\over2}(\eta^{\beta\sigma}\eta^{\rho\xi}
+\eta^{\beta\xi}\eta^{\rho\sigma})-{1\over3}\eta^{\beta\rho}\eta^{\sigma\xi}
\end{eqnarray}
with $\eta^{\beta\rho}=-g^{\beta\rho}+Q^\beta Q^\rho/m^2_{a_2}$.

In the Regge framework the predictions for the physical
observables are strongly dependent on the phase and sensitive to
the intercept of the trajectory as well. Therefore, to use the
right phase and trajectory is of importance, though there is no
rigorous theory for this purpose. One feasible scheme for  this is
the addition of the phases of the EXD pairs $\pi$-$b_1$ and
$\rho$-$a_2$ as discussed in Ref. \cite{guidal}.

From the $G$-parity of the photon-meson coupling vertex and  the
$\pi N\Delta$ coupling constants related to the isospin,
\begin{eqnarray}\label{mbv}
f_{\pi^- p\Delta^{++}}=-f_{\pi^+ n\Delta^-}=-\sqrt{3}f_{\pi^+
p\Delta^0}=\sqrt{3}f_{\pi^- n\Delta^+}\,,\hspace{0.5cm}
\end{eqnarray}
the production amplitude is written symbolically as
\begin{eqnarray}\label{exd-phase}
{\cal M}=\left\{\begin{array}{c}1\\1/\sqrt{3}
\end{array}\right\}\left[\left(\pm\pi+b_1\right)+\left(\rho\pm
a_2\right)\right],
\end{eqnarray}
for the $\gamma n\to\pi^+\Delta^-$, $\gamma p\to\pi^-\Delta^{++}$
(upper), and $\gamma p\to\pi^+\Delta^0$, $\gamma
n\to\pi^-\Delta^+$ processes (lower), respectively. The phases of
the exchanged meson, thus determined, are the same as those given
in the $\gamma N\to \pi^\pm N$ process \cite{bgyu-pi,guidal}.

\subsection*{Results and discussion}

\begin{figure}[]
    \centering
    \includegraphics[width=0.95\hsize,angle=0]{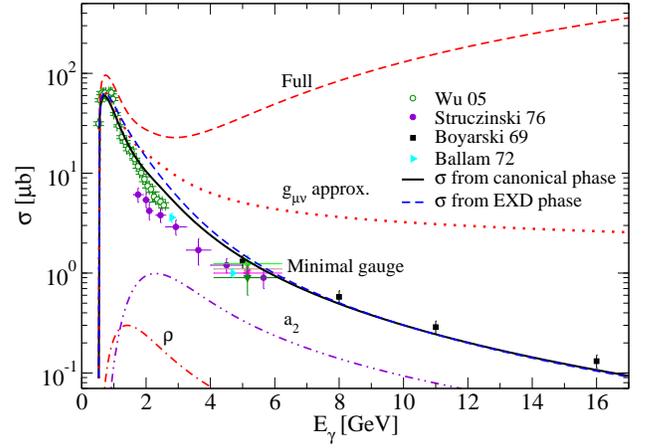}\\
\vspace{0cm}
    \caption{
    Total cross section for $\gamma p \to\pi^- \Delta^{++}$ from $\pi+\rho+a_2$ exchanges.
    Red dashed line is from the $\Delta$-pole with the full propagation in Eq. (\ref{uch1})
    and the canonical phase for $\pi$.
    Red dotted line is the case of the approximation  $\Pi^{\mu\nu}_\Delta\approx
    -g^{\mu\nu}$ for the $\Delta$-propagation in Eq. (\ref{amp1-u}).
    The convergent cross sections shown by the solid and blue dashed lines result
    from the minimal gauge prescription with the phase of the $\pi$ exchange denoted
    in the figure legends. Shown in the dash-dotted and dash-dot-dotted
    lines are the respective contributions of the $\rho$ and $a_2$
    exchanges.
    Data are taken from Refs. \cite{Boyarski:1969dw,wu2005,ballam:72,struczinski:76}.
    Data points at $E_\gamma=5$, 8, 11, and 16 GeV are marked by integrating out
    the data of differential cross sections in Ref. \cite{Boyarski:1969dw}.} \label{fig2}
\end{figure}

Given the trajectories for $\pi$, $\rho$, and $a_2$ Regge-poles as
\begin{eqnarray}\label{regge2}
&&\alpha_\pi(t)=0.7\,(t-m_\pi^2)\,,\\
&&\alpha_{\rho}(t)=0.8\,t+0.55 \,,\\
&&\alpha_{a_2}(t)=0.85\,(t-m^2_{a_2})+2 \,,
\end{eqnarray}
respectively, we calculate  the total cross section for $\gamma
p\to\pi^-\Delta^{++}$ and present the result in Fig. \ref{fig2},
where the blue dashed line denotes the total cross section
$\sigma$ from the case of the $\pi$ exchange with the EXD phase,
$1$,  as determined in Eq. (\ref{exd-phase}), and the solid line
is from the case of the canonical phase for the $\pi$ exchange,
respectively.

\begin{table}[]
\caption{\label{cc1} Meson-baryon coupling constants for
$^{(a)}\gamma p\to \pi^- {\Delta^{++}}$ and $^{(b)}\gamma p\to
\pi^+ {\Delta^0}$ processes. The superscript indicates the phase
taken for the process denoted. }
\begin{tabular}{clccc}
                       &Coupling const.              & $^{(a)}$Phase & $^{(b)}$Phase\\
\hline\hline
$\pi$                  &$f_{\pi N\Delta}=2.0$        &${1\over2}(1+e^{-i\pi\alpha_\pi(t)})$& ${1\over2}(1+e^{-i\pi\alpha_\pi(t)})$  \\%
\hline
$\rho$                 &$g_{\gamma\pi^\pm\rho}=0.224$&$1$&$e^{-i\pi\alpha_\rho(t)}$  \\%
                       &$f_{\rho N\Delta}=8.57$      &      &    \\%
\hline
$a_2$                  &$g_{\gamma\pi a_2}=-0.276$   &$1$&$e^{-i\pi\alpha_{a_2}(t)}$   \\%
                       &${f_{a_2N\Delta}\over m_{a_2}}=-3{f_{\rho N\Delta}\over m_\rho}$&           &  \\%
\hline\hline
\bigskip
\end{tabular}\label{tb1}
\end{table}

For a better agreement with data as shown in the resonance region
we choose the canonical phase for the dominating $\pi$ exchange in
the absence of $b_1$ exchange in the present calculation.

A summary of the coupling constants and the phases of the
exchanged mesons used for the present calculation is given in
Table \ref{tb1}.

Figure \ref{fig2} shows the convergence of the total cross section
up to $E_\gamma=16$ GeV depending on the treatment of the
$\Delta$-pole, i.e., by using Eq. (\ref{pi-regge}) for the full
propagation of $\Delta$, or by the minimal gauge as in Eq.
(\ref{min}). We also examine the validity of the approximation,
$\Pi^{\mu\nu}_\Delta\approx -g^{\mu\nu}$ widely used for hadron
reactions involved in the $\Delta$ coupling. Both the cases of the
$\Delta$-pole with the full propagation (red dashed line) and with
the approximation of $\Pi_\Delta^{\mu\nu}\approx -g^{\mu\nu}$ in
Eq. (\ref{propa2}) (red dotted line) are highly divergent as the
photon energy increases, whereas a good behavior of the cross
section for convergence is obtained by using the minimal gauge as
shown by  the solid and the blue dashed lines. These findings
confirm the validity of the minimal gauge for the high energy
behavior of the $\pi\Delta$ photoproduction.

\begin{figure}[]
    \includegraphics[width=0.95\hsize,angle=0]{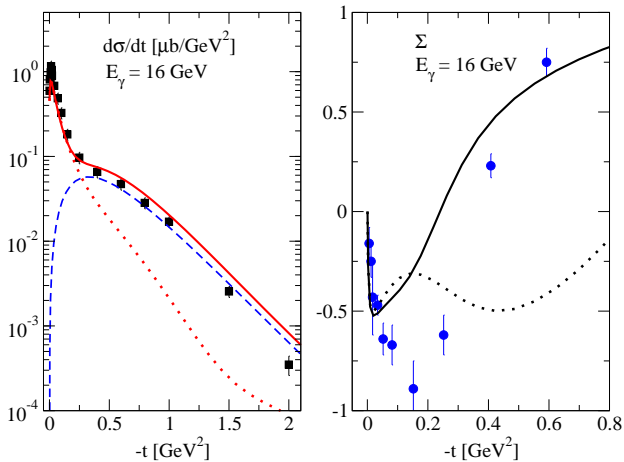}
    \caption{Differential cross section and
    photon polarization asymmetry for $\gamma p\to\pi^-\Delta^{++}$ at $E_\gamma=16$ GeV.
    The solid and dotted lines are from the $\rho+\pi$ exchanges in
    the minimal gauge with and without tensor meson $a_2$, respectively.
    The dashed line in the left panel depicts the $a_2$ contribution.
    Data are from Refs. \cite{quinn79,Boyarski:1969dw}. }
    \label{fig3}
    \vspace{0.5cm}
    \bigskip
\end{figure}

\begin{figure}[]
    \includegraphics[width=0.95\hsize,angle=0]{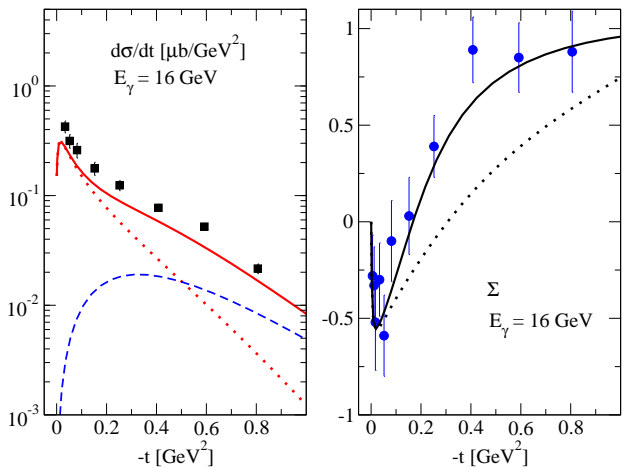}
    \caption{ Differential cross section and photon polarization asymmetry
    for $\gamma p\to\pi^+\Delta^0$ at $E_\gamma=16$ GeV.
    Notations for the curves are the same as in Fig. \ref{fig3}.
    Data are from Refs. \cite{quinn79,Boyarski:1969dw}.
    }
    \label{fig4}
\end{figure}

The role of the tensor meson $a_2$ in the reaction process is
evident in the differential cross section $d\sigma/dt$ and the
photon polarization asymmetry which is defined as
\begin{eqnarray}
\label{sd0} \Sigma &=& \frac{d\sigma^\bot-d\sigma^\parallel}{
d\sigma^\bot+d\sigma^\parallel}\ ,
\end{eqnarray}
in the c. m. frame of the pion production plane.

We present the cross sections for the differential and the photon
polarization asymmetry from the SLAC data for the $\gamma
p\to\pi^-\Delta^{++}$ at $E_\gamma=16$ GeV in Fig. \ref{fig3}, and
for the $\gamma p\to\pi^+\Delta^{0}$ in  Fig. \ref{fig4},
respectively. While our model predictions are remarkably in good
agreement with data, the tensor meson $a_2$ exchange gives the
contribution crucial to agree with existing data in comparison to
the dotted lines resulting from the $\rho+\pi$ exchanges without
$a_2$ in the minimal gauge.

To summarize, we have investigated the $\gamma
p\to\pi^-\Delta^{++}$ process up to $E_\gamma=16$ GeV with our
focus  on the production mechanism of the reaction process at high
energy. For this purpose we constructed a  Born term model where
the $\pi$ exchange is reggeized in the $t$-channel with the
$u$-channel $\Delta$-pole included for gauge invariance in
addition to the $s$-channel proton-pole and the contact term. In
order to make convergent the energy-dependence of the reaction
cross section against the divergence of the $\Delta$-pole at high
energy we utilized the minimal gauge prescription to simplify the
$\Delta$-pole, which  is possible for the antisymmetric form of
the charge coupling terms in the $\gamma\Delta\Delta$ vertex due
to the Ward identity at the vertex.

We further showed that the tensor meson $a_2$ exchange plays the
key role to agree with the existing data on the differential and
photon polarization asymmetry for $\gamma p\to\pi^-\Delta^{++}$
and $\gamma p\to\pi^+\Delta^{0}$  at high energy. For doing this
we constructed a new effective Lagrangian for the tensor
meson-nucleon-$\Delta$ coupling in this work and demonstrate its
validity by obtaining quite improved differential cross section as
well as photon polarization asymmetry with the coupling constant
${f_{a_2N\Delta}\over m_{a_2}}=-3{f_{\rho N\Delta}\over m_\rho}$
which is deduced from the duality plus vector dominance.

While revealing the well-known approximation
$\Pi_{\Delta}^{\mu\nu}\approx -g^{\mu\nu}$ to be valid only in the
limited energy region near threshold, we understand the production
mechanism in this minimal gauge as the dominance of the $\pi$
exchange incorporating with the $a_2$ exchange rather than the
$\rho$.

\section*{Acknowledgment}
This work was supported by 2012 Korea Aerospace University Faculty
Research Grant No. 2012-01-015.


\end{document}